# A combined theoretical and experimental study of the influence of different anion ratios on lithium ion dynamics in ionic liquids

Volker Lesch[1], Sebastian Jeremias[1], Arianna Moretti[1], Stefano Passerini[1,2,*], Andreas Heuer[1],* Oleg Borodin[3]

1 Institute of Physical Chemistry, Westfälische Wilhelms-Universität Münster, Corrensstrasse 28/30, 48149 Muenster, Germany

2 Helmholtz Institute Ulm, Karlsruhe Institute of Technology, Albert Einstein Allee 11, 89081 Ulm, Germany

3. U. S. Army Research Laboratory, Electrochemistry Branch, Sensors & Electron Devices Directorate, 2800 Powder Mill Rd.; Adelphi, MD 20783

## Abstract

In this paper we have investigated via experimental and simulations techniques the transport properties, in terms of total ionic conductivity and ion diffusion coefficients, of ionic liquids and lithium salt mixtures composed of two anions, bis(fluorosulfonyl)imide (FSI) and bis(trifluoromethanesulfonyl)imide (TFSI), and two cations, N-ethyl-N-methylimidazolium (emim) and lithium. The comparison of the experimental results with the simulations shows an exceptional agreement over a wide temperature range. Of particular interest is the high agreement achieved at lower temperature, which is a result of the extended simulation length of, at least, a factor of 5 compared with previous work. The results show as the addition of TFSI favor the formation of lithium dimers ($Li^+ – TFSI^- – Li^+$). A closer analysis of such dimers shows that involved lithium ions move nearly as fast as single lithium ions although they have a different coordination and much slower TFSI exchange rates.

Corresponding Authors: **andheuer@uni-muenster.de**, stefano.passerini@kit.edu

# 1 Introduction

Within the last decades the use of ionic liquids (ILs) in different fields has been established. Because of their unique properties ionic liquids are promising materials for different applications, e.g., as solvent in synthesis, as catalyst and in electrochemical applications [1 - 3]. The bulky, organic cation can be easily tailored by selecting different side chains to specific demands [2, 3]. Especially for their application in electrochemical devices, ILs offer promising physical properties like high thermal stability, low vapor pressure and wide electrochemical stability window [3, 4, 5 ].

However, due to the low ambient conductivity of IL/Li-salt mixtures compared to conventional electrolytes, improvements related to this property are of main interest [5]. Several attempts have already been made with respect to this issue. For example, Kühnel et al. added organic electrolytes to IL/LiTFSI mixtures enhancing conductivity, thermal stability and suppressing aluminum corrosion [6]. Furthermore, mixtures of $pyr_{13}$FSI (N-methyl-N-propylpyrrolidinium bis(fluorosulfonyl)imide) and $pyr_{14}$TFSI (N-butyl-N-methylpyrrolidinium bis(trifluoromethanesulfonyl)imide) show higher conductivities and lower melting points due to ion mismatch [7,8]. However, only a microscopic understanding of fundamental transport processes would make it possible developing strategies to increase the ionic conductivity. The big challenge is to successfully correlate experimental and theoretical results.

Molecular dynamics (MD)-simulations is a powerful tool to elucidate microscopic mechanisms of complex molecular systems. Accurate representation of intermolecular interactions is important for predicting structural and, especially, dynamic properties of ILs doped with lithium salts. Due to strong polarization of anions by a small $Li^+$ cation, it is important to include many-body polarizable terms in the intermolecular potential (force field) used in MD simulations. Previous simulations using atomic polarizable force field for liquids, electrolytes and polymer (APPLE&P) have accurately predicted both structural and transport properties of pure ILs and IL doped with $Li^+$ salts [9-13].

Previous MD simulations with APPLE&P focused on the pyr-based ILs doped with lithium salts. [9,10,11]. It has been shown that the coordination does not vary when changing the cation from $pyr_{13}$ to $pyr_{14}$. Lithium aggregation sharing TFSI anions as bridges were also observed but the simulations at lower temperatures are too short to break up this aggregation[13]. Two mechanisms for lithium diffusion was observed. One way is the structure diffusion, which means that lithium ions move by exchanging its first coordination shell. The other way is the vehicular mechanism, which means lithium ions move with their first coordination shell. This was investigated in more detail by adding an additional potential function, which stabilize the coordination of TFSI on lithium ions, resulting to slowing down the TFSI exchange out of the first coordination shell and thus, slowing lithium ion diffusion. Thus, the authors suggested that the Li-TSFI aggregate move much slower, giving rise to a slower vehicular mechanism as compared to the structure diffusion mechanism. Both transport possibilities were investigated by Li et al. as a function of lithium ion concentration [10,13]. They found an increasing importance of the vehicular mechanism by decreasing the amount of lithium ions and decreasing the temperature since the mean square displacement of the lithium ions on the time scale of the coordination shell lifetime somewhat increases.

Solano et al. used the correlation between experimental and numerical information to characterize the systems $pyr_{14}TFSI$ and $pyr_{14}FSI$[11]. Due to the relatively short trajectories of 3 ns used in that work, simulations were restricted to high temperatures and no direct overlap with the experimentally accessible regime was possible.

In this work we investigate systems with different ratios of FSI and TFSI anions, which are promising candidates for electrochemical applications. As cation we use N-ethyl-N-methylimidazolium (emim) This work is motivated by the experience from organic electrolytes in lithium-ion batteries. In fact, mixtures of different carbonates display improved properties like SEI formation and lowering of the melting point and viscosity [14]. It is known that FSI has a film forming ability on the anode while TFSI has a higher thermal stability. Thus, also for ILs the use of anion mixtures may be useful. Although FSI and TFSI are structurally quite similar their properties are surprisingly different, so, their effects on each other are very interesting. Among other things we explicitly study whether the presence of Li-anion clusters indeed is a viable mechanism to slow down the lithium diffusivity as suggested in previous work [13].

## 2  Experimental setup

The ionic liquids and LiFSI were purchased from Solvionic while LiTFSI was purchased from 3M. All compounds were individually dried under vacuum at room temperature for 24 hours with a turbomolecular pump ($10^{-7}$ bar). The four compounds were mixed to make several mixtures, which compositions are reported in Table 1. Each mixture was further dried at room temperature for 24 hours with a vacuum oil pump ($10^{-3}$ bar) and for additionally 24 hours with a turbomolecular pump. Unless otherwise stated, all sample preparation was carried out in a dry room (dew point < -50°C).

**Pulsed field gradient-NMR (PFG-NMR)**

The investigated mixtures were flame sealed in NMR tubes inside the dry room to prevent any contamination. To determine diffusion coefficients of different ion species we have used a Bruker-NMR-spectrometer (Germany) with permanent field strength of 4.7 T. The diffusion probe head was a "Diff30" (Bruker, Germany) with selective rf. inserts for $^1$H, $^{19}$F and $^7$Li. The maximum gradient strength was technically limited to 1.8 T/m. We changed temperature with the water-cooling unit stepwise from 10 °C up to 49 °C. Higher temperatures (60 °C and 70 °C) were reached by additional heated air flow (500 L/h). For the PFG-NMR experiments we have used the STE- and the dstegp3s sequences according to the Bruker library (Top Spin 3.0). The dstegp3s sequence is necessary in case of air flow heating to suppress convection artifacts. The echo attenuation was evaluated by a monoexponential fitting.

**Ionic conductivity measurements**

The temperature dependence of the mixtures conductivity was measured by mean of an automated conductivitymeter ("MaterialMates Italia") equipped with a frequency analyzer and a thermostatic bath. The mixturess were loaded in sealed conductivity cells (inside the dry room) containing two platinum electrodes previously calibrated using a 0.01M KCl aqueous solution. The cells were cooled down to -40°C for 18 hours before starting to increase the temperature in 2 °C steps. At each temperature, the samples were left to equilibrate for 1hour.

## 3  Simulation methodology

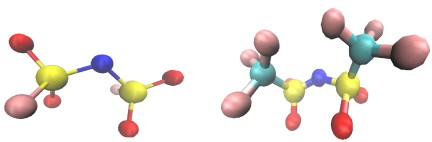

Figure 1: Structure of FSI.  Figure 2: Structure of TFSI.

(blue: N; yellow: S; red:O;cyan:C; pink:F)

All simulations were performed with the MD-simulation package AMBER 10 [15]. This software was extended by a Buckingham Potential and a Thole screening. These two modifications make it possible to use the many body polarizable force field APPLE&P [9,16].

The starting structure was produced randomly in the gas phase. First, the simulation cell was shrunk. After this an equilibration run of 10 ns in the NpT ensemble at 1 bar was performed to obtain a homogeneous system. The averaged density of this step was used for the followed production run in the NVT ensemble. The comparison of experimental and estimated densities is shown in Tab. 2. The `Berendsen` thermostat was used to control the temperature, the `SHAKE` algorithm to constraint the bonds containing hydrogen. The elementary integration step was 1 fs. The electrostatic interactions were calculated via the particle-mesh `Ewald` summation (Size of grid:48x45x45, interpolation order: 4, Ewald coefficient: 0.27511, Switch function for direct Coulomb: cubic). The simulation box contained 180 emim ions and 17 lithium ions as cations and (0|197; 17|180) FSI/ TFSI anions. Only for the system with a ratio of 1:1 FSI:TFSI we used 208 emim ,20 lithium ions and (114|114) FSI/ TFSI anions.

Table 1: Composition (in molecules) of the five investigated mixtures. The same number of ions was considered in the simulation boxes.

| TFSI [%] | 100 | 91 | 50 | 9 | 0 |
|---|---|---|---|---|---|
| emim | 180 | 180 | 208 | 180 | 180 |
| TFSI | 197 | 180 | 114 | 17 | 0 |
| FSI | 0 | 17 | 114 | 180 | 197 |
| Li | 17 | 17 | 20 | 17 | 17 |

The simulation temperature was varied between 298 K and 403 K. Our trajectories were sufficiently long to determine the diffusivities in a reliable manner (e. g. 150 ns for T = 333 K). Compared to previous works we extend the trajectories of more than a factor of 5 [11, 10]. The detailed microscopic analysis was performed at 333 K where we can directly compare with the experimental data. In the following the systems are characterized by the anion fraction of TFSI, i.e. $TFSI_0$, $TFSI_9$, $TFSI_{50}$, $TFSI_{91}$ and $TFSI_{100}$ (see Table 1).

Table 2: Comparison of experimental and simulation densities. The experimentally determined viscosity is also reported. All values refer to 333 K.

| TFSI [%] | | 100 | 91 | 50 | 9 | 0 |
|---|---|---|---|---|---|---|
| $\varrho$ [ g/cm³ ] | Exp. | 1.5072 | 1.5021 | 1.4813 | 1.4422 | 1.435 |
| | Sim. | 1.50 | 1.50 | 1.47 | 1.43 | 1.42 |
| $\eta$ [ mPas ] | Exp. | 16.188 | 14.1595 | 10.482 | 8.781 | 9.5805 |

## 4 Diffusion coefficients

The self-diffusion coefficient of each ionic species was determined via PFG-NMR experiments and simulations, using the Einstein-Smuchlowski equation [17].

$$D = \lim_{t \to \infty} \frac{MSD_{ion}}{6t} \quad (1)$$

In eq. 1 $MSD_{ion}$ is the mean-square displacement of the specific ion, <> denotes the ensemble average and t is the time. For the comparison of experiment and simulation the data from the simulations were corrected for hydrodynamic effects, which show up finite size effects. It is given by

$$\Delta D_{FSC} = \frac{2.837 k_B T}{6 \pi \eta L} \tag{2}$$

where $k_B$ is the Boltzmann constant, $T$ is the temperature, $\eta$ is the viscosity and $L$ is the box length [18]. The viscosity values were experimentally determined (see Table 2). The temperature ranges of simulation and experiment do overlap between 298 K and 333 K. For low temperatures sufficient equilibration of the simulated configuration is necessary. PFG-experiments are limited to a maximum temperature of 343 K due to technical reasons. In practice, we have fitted the experimental data by the VTF relation in order to have a complete overlap with the experimental data. The VTF fit function is given by

$$D = D_0 \cdot e^{-\frac{B}{T-T_0}} \tag{3}$$

Here $D_0$ and $B$ are material dependent constants, which can be freely adjusted. $T_0$ is hold constant for all fits to make the results comparable to each other. The fit parameters can be compared to experimental data from Tokuda et al. [19]. They investigated the ionic liquid without lithium salt and found nearly the same values for $B$ and $T_0$. The fitting parameters for our results, displayed in Fig. 3, are shown in tab. 3. For further information see supplementary information.

Table 3: Fit parameter for the VTF fit of diffusion and conductivity data.

|  | $D_0$ [$10^{-10}$ m$^2$/s] | B [K] | $T_0$ [K] |
| --- | --- | --- | --- |
| $D_{emim}$ | 9e-9 ± 7e-10 | 761 ± 13 | 157 |
| $D_{TFSI}$ | 8.9e-9 ± 5e-9 | 858 ± 106 | 157 |
| $D_{Li}$ | 8.9e-9 ± 8e-10 | 934 ± 17 | 157 |
|  | $\sigma_0$ [mS/cm] | B [K] | $T_0$ [K] |
| $\sigma$ | 791 ± 43 | 684 ± 9 | 157 |

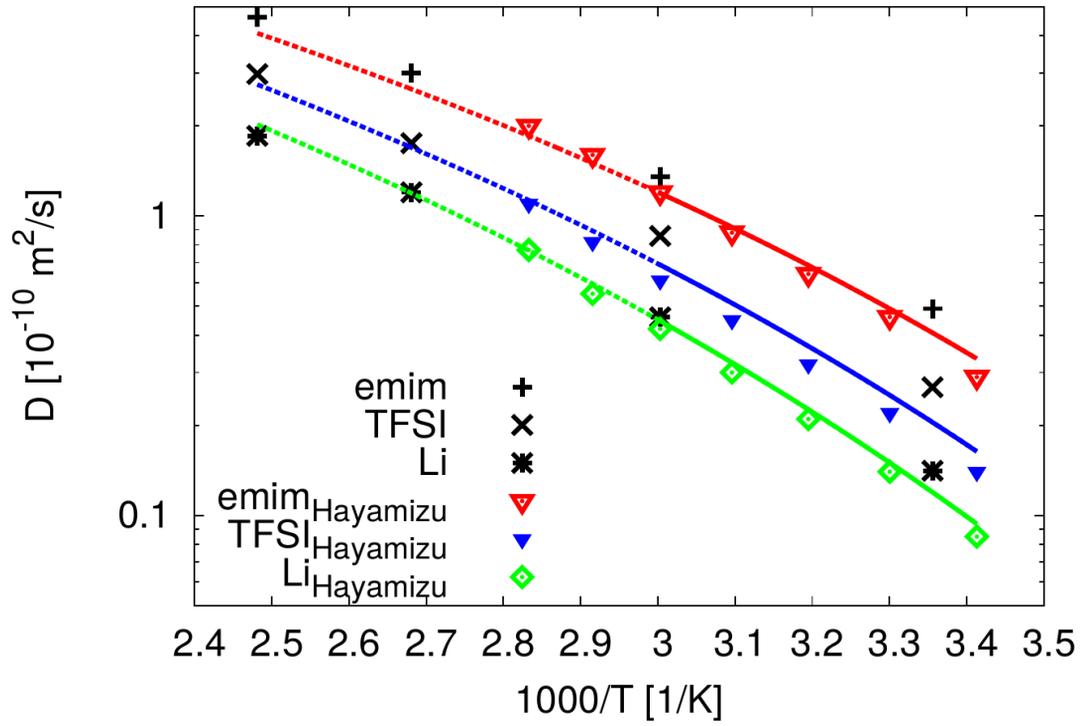

Figure 3: Diffusion coefficients for the system TFSI$_{100}$; lines represent the VTF fit to the experimental data (solid lines: direct overlap; dashed lines: extrapolation of exp. data), the data points the simulation results and the diffusion coefficients determined by Hayamizu et al.[20].

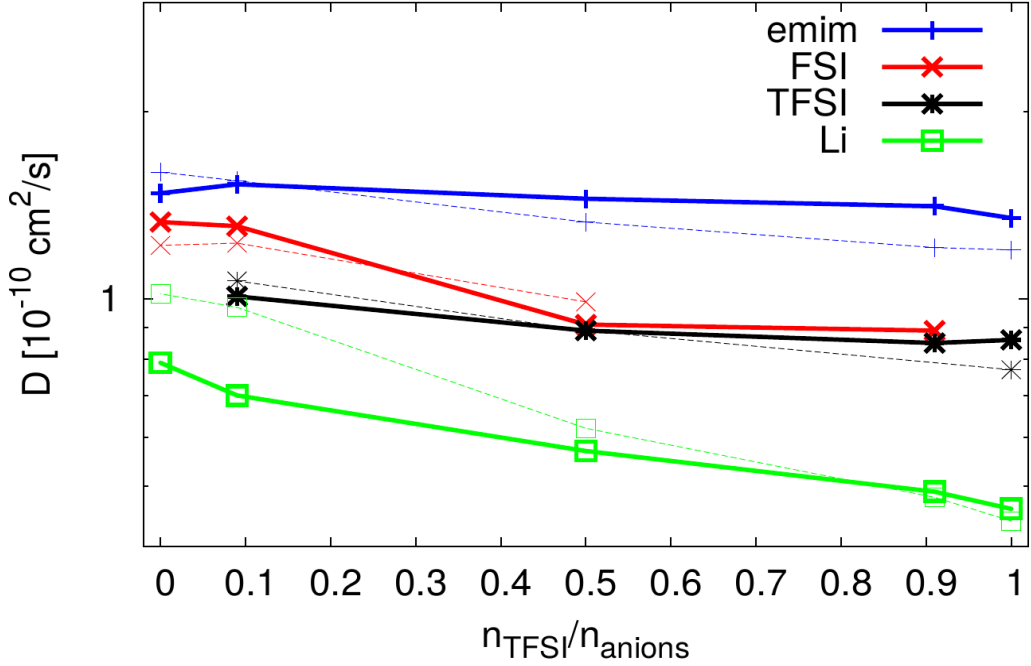

Figure 4: The diffusion coefficients in dependence on the TFSI fraction for T = 333 K; the dashed lines connect the experimental data, the solid lines represent the simulated data.

Fig. 3 and 4 compare the experimental and simulation data. For the complete temperature range a good agreement between experiment and simulation can be observed (see Fig. 3). Furthermore, we compared our data with PFG-NMR experiments from Hayamizu et al. and found also a very good agreement [20]. We checked that the small deviations are not due to insufficient sampling. In Fig. 4 the dependence of each ionic species on the anion composition is shown. The agreement between experiment and simulation is again very promising. Significant deviations (approx. 20%) are only visible for the lithium ion diffusivity, which experimental values displays a stronger dependence on TFSI concentration than the simulation results. This significant deviation for lithium ion diffusivity can be observed over the complete temperature range(see supplementary material). The difference between $TFSI_0$ and $TFSI_{100}$ for the anion and lithium ion diffusion shows the same order as reported by Solano et al. for systems with the cation $pyr_{14}$ [11]. They also found that the anion has only a small influence on the diffusion of

pyr$_{14}$. This result is in accordance with our data because the emim diffusion constant is nearly constant as shown in Fig. 4.

## 5 Conductivity

The Einstein relation

$$\sigma = \lim_{t \to \infty} \frac{e^2}{6tVk_BT} \sum_{i,j} z_i z_j \langle R_i(t) - R_i(0) \rangle \langle R_j(t) - R_j(0) \rangle \tag{4}$$

can be used to calculate the ionic conductivity from MD trajectories. In eq. 4 $e$ is the elementary charge, t is the time, V is the simulation box volume, T is the temperature and $z_{i,j}$ are the charges of the ions i, j. Due to the cross terms in the sum this value can only be determined with poor statistics compared to ion self-diffusion coefficients. The conductivity can be split into a part without cross terms, reflecting the uncorrelated part. It reads

$$\sigma_{uncorr} = \lim_{t \to \infty} \frac{e^2}{6tVk_BT} \sum_{i,j} z_i^2 \langle R_i(t) - R_i(0) \rangle^2 = \frac{e^2}{Vk_BT} \sum_i n_i D_i \tag{5}$$

where n$_i$ is number of the ionic specie and D$_i$ is the connected diffusion coefficient. The degree of uncorrelated motion α can be defined as the ratio between eq. 4 and eq. 5, i.e.

$$\alpha = \frac{\sigma}{\sigma_{uncorr}} \tag{6}$$

α = 1 means complete uncorrelated motion, while α=0 stands for a complete correlated motion so all cations move together with anions. Due to the poor statistic for the long time limit of the cross terms α is determined in the sub-diffusive regime by utilizing the first 2-5% of the trajectory [21]. Therefore, we split our trajectory and investigate in which time interval α can be determined with good statistics. We calculated α in the regime from 0.4 ns to 1 ns (corresponding to less than 2% of the total simulation) because, above an interval of 2 ns, α starts to show statistical errors (see supplementary material).

Table 4: Averaged α values from experiment and simulation.

| T [K] | Sim. | Exp. |
|---|---|---|
| 298 | 0.66 | 0.63 |
| 333 | 0.73 | 0.64 |

In tab. 4 the averaged values of α are shown. The value of α was independent from the anion composition for the simulations as well as for the experiment. Due to the uncertainty for α of 0.1 [10,19], the agreement between experiment and simulation is very good.

The "ideal" conductivity (see eq. 5) can be calculated from the diffusion coefficients. These were determined in the long time limit like described in the previous paragraph. Under the assumption that α(t) is time-independent for even longer times we can invert eq.6 in order to obtain the conductivity from $\sigma_{uncorr}$.

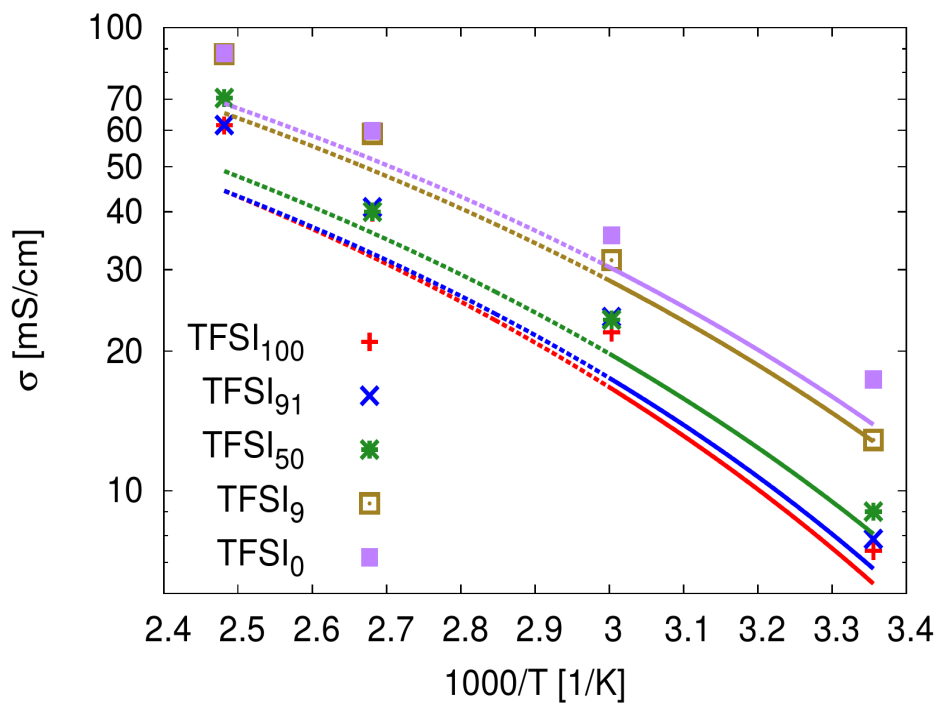

Figure 5: The conductivity for the experimental data (solid lines) and simulation results (data points).

In Fig. 5 a comparison of the conductivities is shown. The determined conductivities from simulations differ at most 30% from the experimental ones and show the same dependence on TFSI concentration. Part of the residual deviations result from the differences in the α-value at the higher temperature. Importantly, the two systems with the lowest content of TFSI show a significantly higher conductivity than the other three systems in experiment as well as in simulation.

**6 Lithium dimers**

In order to establish the relationship between transport and structural properties, the structure of the lithium coordination shell and ionic aggregated has been examined.

We start with the analysis of the radial distribution function (RDF) of the lithium ions, shown in Fig. 6. For all systems the first coordination sphere is well defined but the systems strongly differ in the amount of lithium ions in the first coordination shell. From 0% to 91% TFSI the relative intensity of the first coordination sphere increases. The $TFSI_{100}$-system behaves similar as the $TFSI_{91}$-system. The strongest effect is present for $TFSI_9$ and $TFSI_0$ where a strong reduction of direct lithium ion neighbors is observed. Compared with the $TFSI_{100}$-system a factor of nearly 3 is observable. The first peak and the second peak are shifted compared to the simulations of Borodin et al.[13]. Thus maybe results out of the different cation or higher lithium ion concentration. The ratio between both coordination shells is in the same order of magnitude.

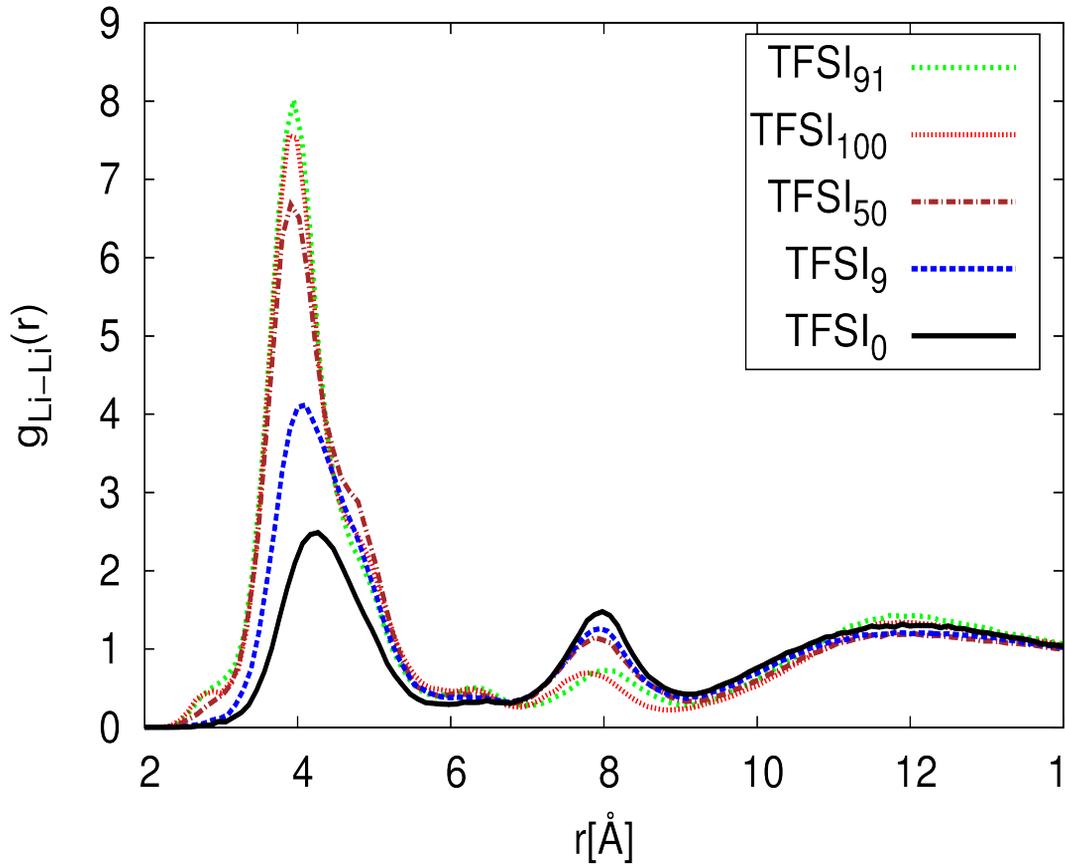

Figure 6: RDF of lithium ions with themselves.

As a compensation, for the system TFSI$_0$ the second coordination shell is much more pronounced. Fig. 7 shows a possible lithium ion dimer. Here, TFSI acts as a bridge because the peak position of the first coordination shell is in accordance with the distance between two oxygen atoms of TFSI. FSI can also act as a bridge but the probability is much lower because TFSI is sterically much more demanding and coordinates stronger.

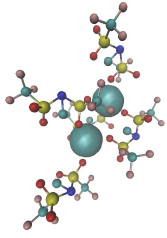

Figure 7: Lithium dimer and TFSI as bridge.

To analyze such dimers we first determine their lifetimes. Moreover, we calculate the exchange rate of anions out of the first coordination shell. For both values an Auto-correlation function (ACF) was used

$$ACF(t) = \frac{\langle B_{ij}(t) B_{ij}(0) \rangle}{\langle B_{ij}(0) B_{ij}(0) \rangle} \qquad (7)$$

where $B_{ij}=1$ if the ions i and j are inside the first coordination shell of each other, corresponding to 5.7 Å (Li-Li) and 2.75 Å (Li-O). Otherwise one gets $B_{ij}=0$. The $\langle \rangle$ denotes the ensemble average. This ACF functions for Li-Li and Li-O are shown in Fig. 8. The decay of the ACF can be fitted via a Kohlrausch-Williams-Watts (KWW) function.

$$ACF(t) = A \cdot e^{-\left(\frac{t}{\tau}\right)^{\beta}} \qquad (8)$$

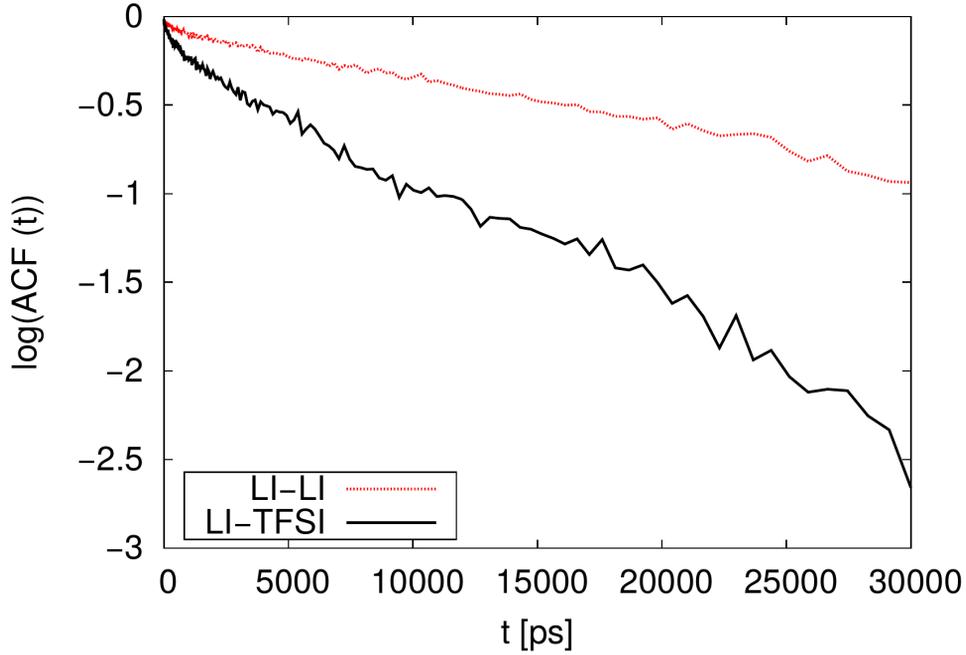

Figure 8: ACF of Li-Li and Li-O$_{anion}$; System: TFSI$_{100}$; T= 333 K.

Based on this fit the lifetimes can be calculated via

$$\tau_{life} = \int_0^\infty e^{-\left(\frac{t}{\tau}\right)^\beta} dt \qquad (9)$$

In Tab. 5 the resulting lifetimes of such dimers are presented. Additionally, the exchange rates of FSI and TFSI inside a lithium ion dimer, on a single lithium ion and percentage of lithium ions, which are involved in such dimers, are shown. The comparison of low and high contents shows an interesting behavior. The addition of FSI to TFSI$_{100}$ does not influence the lifetime of the dimers. However, an increase of the lifetime by a factor of two is observed if TFSI$_0$ and TFSI$_9$ are compared, i.e., upon addition of TFSI. The exchange rates of the anions are comparable to those determined by Li et al. [10]. These authors sum over all lithium ions without a distinction between dimers and single lithium ions, obtaining exchange rates of the same order of magnitude. Borodin et al.[13] determine a life time of 7.1 ns for such dimers at T=393 K using a slightly different force field. Compared to the life times at T=373 K (roughly 11 ns) and T=403 K (5 ns) of our simulations both are in good agreement considering a difference in the force field.

Table 5: Lifetimes and percentage of lithium ions inside a dimer and lifetime of FSI/TFSI in the first coordination shell of a single lithium ion (all times are in ns)

| System [% TFSI] | Li-Li | $Li_{dimer}$ - FSI | $Li_{dimer}$ - TFSI | $Li_{single}$ - FSI | $Li_{single}$ - TFSI | $(\# Li_{dimer})/(\# Li_{tot})$ |
|---|---|---|---|---|---|---|
| 100 | 32.6 | - | 9.7 | - | 2.0 | 36.3 |
| 91 | 32.9 | 8.8 | 13.1 | 0.75 | 1.6 | 35.3 |
| 50 | 29.5 | 5.4 | 11.8 | 0.57 | 1.5 | 40.1 |
| 9 | 7.9 | 1.7 | 6.7 | 0.49 | 1.1 | 31.0 |
| 0 | 4.3 | 0.9 | - | 0.48 | - | 18.8 |

The system without TFSI is noticeable because less than 20% of all lithium ions are involved in dimers. A small amount of TFSI increases this fraction to over 30%, indicating that TFSI strongly stabilizes lithium dimers. Due to the long life times of the dimers and the involved amount of lithium ions one may wonder about the lithium ion transport in presence of the dimers. We explicitly determined the MSD for different subensembles of particles. In this way we can proceed without the addition of an external perturbation done in ref.[13]. The MSD of the single lithium ions with TFSI exchange were compared with those lithium ions for which no TFSI exchange was observed in a given time window. Furthermore, the MSD for lithium ions inside a dimer were determined in the time interval in which the dimer is stable. The lifetimes at single lithium ions show a longer lifetime for TFSI inside the first coordination shell which is in accordance with the radial distribution of Li-$O_{anion}$ because TFSI is much more favored as coordination partner (see supplementary material).

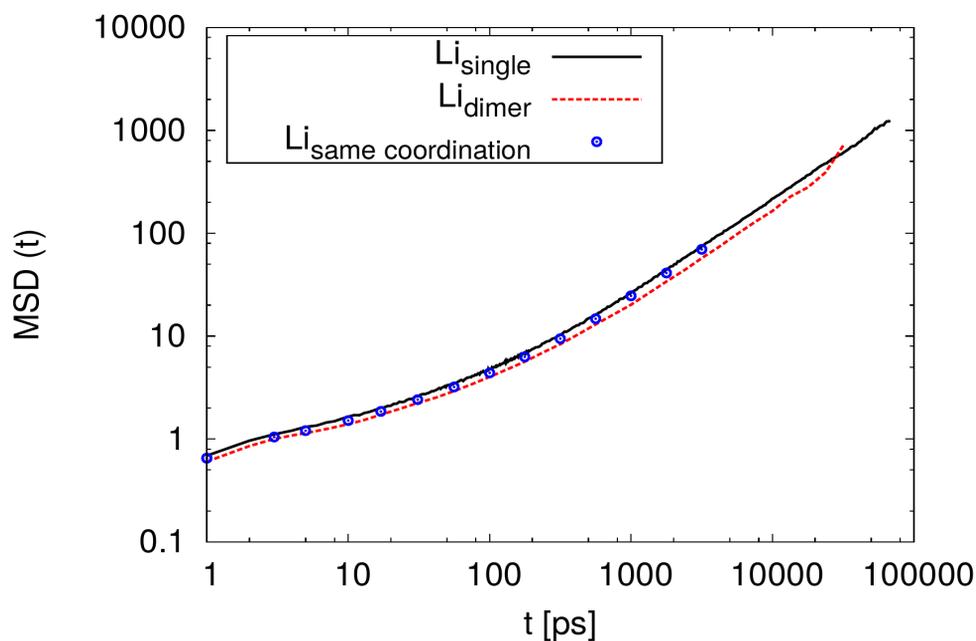

Figure 9: Comparison of MSD of lithium ions in dimers and single ions at T = 333K

Fig. 9 compares the MSD of dimerized and single lithium ions for the system $TFSI_{100}$. Moreover, Fig. 9 reports the MSD of single lithium ions not exchanging their first coordination shell for 7 ns. It is clearly shown, that the TFSI exchange does not influence the lithium ion diffusion. More specifically, the diffusivity of both lithium ion species just differs by 10% at the most. Of course, for $t > \tau_{Li-Li}$ both MSD curves have to merge because the local coordination of a lithium becomes uncorrelated to its initial configuration.

These results suggests that the previous observation [13] concerning the slowing down of the lithium ions may be related to the addition of the binding potential, giving rise to a significantly modified local structure, rather than to the corresponding increase of the life time of the first neighbor shell. Interestingly, as reported in Ref. [13] this addition hardly modified the diffusivity of the large cation. This is consistent with our results, reported in Fig. 4, showing that the emim diffusivity is independent from the anion composition and in this sense somewhat insensitive to the local structure around the anion.

## 7 Conclusions

In this paper we have investigated the transport properties, total ionic conductivity and ion diffusion coefficients of ionic liquids and lithium salt mixtures composed of two anions, bis(fluorosulfonyl)imide (FSI) and bis(trifluoromethanesulfonyl)imide (TFSI), and two cations, N-ethyl-N-methylimidazolium (emim) and lithium, via experimental and simulation techniques. The comparison of the results displays a very good agreement over a wide range of temperatures as well as different FSI/TFSI ratios. The agreement at lower temperature is especially remarkable because no direct overlap with such long trajectories of simulation and experimental results have been reported so far [9-11,16]. Based on this promising agreement between simulations and experiment (also with the data from Ref. [20]) one may explore the microscopic information of the simulation data in order to elucidate some of the microscopic mechanisms, present in these systems. Apart from the excellent agreement between experiment and simulation we can observe the same trends like Borodin et al. related to diffusive properties and life time of lithium dimers for higher temperatures. Therefore, we extend the trajectories at lower temperatures to achieve good statistics for the comparison with the experiment. Here, we have studied in particular the properties of the lithium dimers with respect to life times, exchange rates and percentage of involved lithium ions. Due to the possibility of performing much longer simulations, temperatures in the experimentally relevant temperature regime are reached and a characterization of lithium dimers is possible. The TFSI anion plays an important role for the lithium dimers. Adding a small amount of TFSI results in the strong increase of the number of lithium dimers and their lifetimes. The dimers are stabilized and the anion exchange rate becomes much slower. However, the lithium dimers are nearly as fast as single lithium ions for the temperature accessible in this present work, this is likely due to the roughly similar size of the lithium dimers and $Li^+(TFSI)_4$ complexes. Similar $Li^+$ diffusion coefficient in the dimers and single Li$^+$ should be contrasted with quite different anion-Li$^+$ exchange rate in them. This observation suggests that anion – Li$^+$ exchange rate does not influence Li$^+$ transport, which is a different conclusion from the one reached in the previous study that showed that application of the biasing potential to significantly decrease the Li$^+$-anion exchange slows down Li$^+$ transport by 30% [13]. One should note, however, that the additional biasing potential also perturbed the IL structure, while keeping the average Li$^+$-TFSI coordination the same.

Acknowledgement

Sebastian Jeremias, Arianna Moretti and Stefano Passerini would like to thank the financial support of BMBF within the project "MEET Hi-END - Materialien und Komponenten für Batterien mit hoher Energiedichte" (Förderkennzeichen: 03X4634A). The work is financially supported by the SafeBatt project from BMBF (Förderkennzeichen: 03X4631A).

**References**

(1) Rogers, R. D.; Seddon, K. R. Science 2003, 302, 792–793.

(2) Seddon, K.; Boghosian, S.; Dracopoulos, V.; Kontoyannis, C.; Voyiatzis, G. Boghosian,

S 1999, 131–135.

(3) Bedrov, D.; Borodin, O.; Li, Z.; Smith, G. D. The Journal of Physical Chemistry B 2010, 114, 4984–4997.

(4) Armand, M.; Endres, F.; MacFarlane, D.; Ohno, H.; Scrosati, B. Nature Materials, 2009, 8, 621–629.

(5) Galinski, M.; Lewandowski, A.; Stepniak, I. Electrochimica Acta 2006, 51, 5567 – 5580.

(6) Kühnel, R.-S.; Böckenfeld, N.; Passerini, S.; Winter, M.; Balducci, A. Electrochimica Acta 2011, 56, 4092 – 4099.

(7) Castiglione, F.; Moreno, M.; Raos, G.; Famulari, A.; Mele, A.; Appetecchi, G. B.; Passerini, S. The Journal of Physical Chemistry B 2009, 113, 10750–10759.

(8) Kunze, M.; Jeong, S.; Appetecchi, G. B.; Schönhoff, M.; Winter, M.; Passerini, S., Electrochimica Acta 2012, 82, 69 – 74.

(9) Borodin, O. The Journal of Physical Chemistry B 2009, 113, 11463–11478.

(10) Li, Z.; Smith, G. D.; Bedrov, D. The Journal of Physical Chemistry B 2012, 116, 12801–12809.

(11) Solano, C. J. F.; Jeremias, S.; Paillard, E.; Beljonne, D.; Lazzaroni, R. The Journal of Chemical Physics 2013, 139.

(12) G.D. Smith, O. Borodin, S.P. Russo, R.J. Rees, A.F. Hollenkamp, Phys. Chem. Chem. Phys., 11 (2009) 9884-9897.

(13) O. Borodin, G.D. Smith, W. Henderson, J. Phys. Chem. B 110 (2006) 16879 -16886.

(14) Zhou, Q.; Henderson, W. A.; Appetecchi, G. B.; Montanino, M.; Passerini, S. The Journal of Physical Chemistry B 2008, 112, 13577–13580, PMID: 18828629.

(15) Amber, http://www.ambermd.org

(16) D.M. Seo, O. Borodin, S.-D. Han, P.D. Boyle, W.A. Henderson, J. Electrochem. Soc, 159 (2012) A1489-A1500.

(17) Einstein, A.; von Smoluchowski, M. Untersuchungen über die Theorie der Brownschen Bewegung; Ostwalds Klassiker der exakten Wissenschaften; Deutsch, 1997.

(18) Dunweg, B.; Kremer, K. The Journal of Chemical Physics 1993, 99, 6983–6997.

(19) Tokuda, H.; Ishii, K.; Susan, M. A. B. H.; Tsuzuki, S.; Hayamizu, K.; Watanabe, M., The Journal of Physical Chemistry B 2006, 110, 2833–2839.

(20) Hayamizu, K.; Tsuzuki, S.; Seki, S.; Umebayashi, Y. The Journal of Chemical Physics, 2011, 135, 084505.

(21) Borodin, O.; Gorecki, W.; Smith, G. D.; Armand, M. The Journal of Physical Chemistry, B 2010, 114, 6786–6798.